\documentclass{article}

\usepackage[english]{babel}

\usepackage[
    backend=biber,
    style=alphabetic,
]{biblatex}
\usepackage[a4paper,top=2cm,bottom=2cm,left=3cm,right=3cm,marginparwidth=1.75cm]{geometry}

\usepackage{amsmath,amssymb}
\usepackage{graphicx}
\usepackage[colorlinks=true, allcolors=blue]{hyperref}

\graphicspath{ {figures/} }

\usepackage{lineno}

\usepackage{authblk}

\author[1, *]{George Datseris}
\author[2]{Johannes Lohmann}
\author[3]{Ois\'{i}n Hamilton}
\author[4]{Jacob Haqq-Misra}
\affil[1]{Department of Mathematics and Statistics, University of Exeter}
\affil[2]{Physics of Ice, Climate, and Earth, Niels Bohr Institute, University of Copenhagen, Denmark}
\affil[3]{Stats NZ, Christchurch, Aotearoa}
\affil[4]{Blue Marble Space, Seattle, Washington, USA}
\affil[*]{g.datseris@exeter.ac.uk}

\title{Alternate states and intermingledness in complex high-dimensional systems}

\begin{document}
\maketitle

\begin{abstract}
Many natural systems posses, and can transition between, multiple alternative states.
For example, a climate ``tipping element'' is a climate component that can transition to an alternative steady state due to an external perturbation such as global warming.
Despite the potential impact, existence of alternate states in realistic, complex simulations (e.g. climate models) remain poorly understood.
Arguably a reason for this is the lack of applicable methodology that explicitly targets finite yet high-dimensional datasets.
In this work we utilize recent progress in computational nonlinear dynamics to formulate a workflow that analyses potentially multi-state simulation data and decides algorithmically what are the alternate states contained within, if any are clearly distinguishable.
The framework undergoes an optimization routine that showcases which observables in the data best differentiate the alternate states, and which ones do not differentiate at all, which could be used to guide monitoring and early-warning for multistable components in climate or ecosystems.
Finally, once the alternate states have been found, we define an indicator called ``intermingledness''. It quantifies differences and similarities between alternate states, as well as for their basins of attraction (if applicable), across various diagnostic variables.
We analyse and present results using three diverse climate datasets: Atlantic ocean circulation, atmospheric midlatitude flow, and habitability of exoplanets.
The method is not exclusive to climatic data, but applicable to a variety of cases, including complex networks such as power grids or biological networks.
We also provide easy-to-use open source code for applying the workflow to new data.
\end{abstract}

\section{Introduction}

Complex systems often display several qualitatively different behaviours depending on the initial configuration of the system or forcings that may be applied to it.
Often this occurs because the underlying dynamics are \emph{multistable}, where for the same equations and parameters governing the system, multiple stable states can coexist.
This phenomenon is ubiquitous in nature and in mathematical models \cite{Feudel2018MultistabilityTippingMathematics}, and while it may underline important functional tasks in some systems such as biological ones, it can have drastic negative consequences in others.
For instance, irreversible changes in the state of large-scale sub-systems of the climate (so-called ``tipping points''~\cite{Lenton2008}) are expected at particular levels of global warming~\cite{ArmstronMckay2024TippingReview}). Another example are power grids, where it is imperative that the system remains in the synchronized state for correct operation~\cite{Hartmann2024}, instead of being perturbed to one of the many unsynchronized states that may be available~\cite{Kim2018}.
Beyond multistability, alternate states can manifest themselves as qualitatively different operational behaviour depending on system parameters, such as habitability classes~\cite{HaqqMisra2022}, or alternative system response to stimulus in biological systems~\cite{Gong2005, Bressloff2017}.

It is therefore important to study alternate states in detail for sufficiently realistic mathematical models.
Substantial theoretical work exists, albeit focused on relatively simple, low dimensional models, see e.g., a recent focus issue and references therein~\cite{Feudel2018MultistabilityTippingMathematics}.
Most real world systems, however, are inherently complex, infinitely- or extremely high-dimensional, and often heterotypic.
Indeed, there has been a surge over the last years on simulations showcasing multistability in spatiotemporal climate models of varying degrees of complexity.
Examples include simulating co-existing vigorous and collapsed states of the Atlantic meridional overturning circulation~\cite{Lohmann2021, Romanou2023, Lohmann2024AMOCMultistability, VWE24, DRI25, Boerner2025}, alternative states of ice covered Earth-like planets, such as snowball or waterbelt~\cite{wolf2017constraints, Brunetti2019, Margazoglou2021, Braun2022, Ragon2022, deitrick2023functionality},  alternative circulation and atmospheric flow patterns~\cite{Herbert2020, Simonnet2021, Sergeev2022, Hamilton2023}, ice sheet modelling~\cite{Andernach2026}, cloud regimes in large eddy simulations~\cite{Schneider2019}, and atmospheric dynamical regimes of exoplanets~\cite{carone2015connecting, haqq2018demarcating, zhang2020atmospheric, komacek2025limited}, among others.

The identification of alternate states in the aforementioned high-dimensional simulations involves substantial subjective judgement:
the authors make a best-judgment decision of when the system has stabilized to alternative steady states, (which we will call \emph{attractors}~\cite{Datseris2022NonlinearDynamicsJulia} in the following), using only a couple of representative simulations.
Beyond the lack of rigour, there are some meaningful limitations with this approach: it isn't reproducible, nor transferable to different parameters let alone different models, which is crucial for model intercomparison studies.

In all fairness, a reason for this situation is that there is a lack of workflows explicitly targetting such datasets. On one hand, clustering algorithms can work well when the data points (total simulations) are sufficiently many and not of extremely high dimensionality. Unfortunately, these properties are not easily satisfied in our applications of interest.
Alternatively, one can turn to dynamical systems methodology and in particular the study of multistability which is of direct relevance.
But so far (\cite[\S 2.3]{Pisarchik2022}) there was no generically applicable methodology, especially so for complex systems.
Recently, novel algorithms to identify and study multistability have been developed~\cite{Datseris2022BasinsAttraction, Datseris2023FrameworkGlobalStability}, that are generically applicable to complex systems.
Even these methods however are limited when it comes to very high dimensional data such as realistic climate models, and are thus not trivially transferable to the context we are targetting.


To address this lack of applicable methodology, in this work we generalize the ``featurize and group'' multistability algorithm introduced in Ref.~\cite{Datseris2023FrameworkGlobalStability}, and construct around it a comprehensive computational workflow that is tailored to identifying and analysing multistability or generally alternate states in complex high dimensional data.
The main principle behind ``featurize and group'' is that all simulations that end up at the same distinguishable alternate state should share some common characteristic features (i.e., statistical descriptors).
Once applied, the workflow provides the following: (1) whether alternative states can be distinguished objectively, and if yes (2) which are the physical observables that best separate the data into different states, (3) which physical observables change little between attractors, (4) how much uncertainty each observable has regarding to which state each simulation corresponds to, and (5) how (2-4) change when a model parameter changes (formally called ``continuation''). To estimate (3, 4) we devised a new quantity which we call \emph{intermingledness}, inspired by established concepts in dynamical systems theory.

In \S\ref{sec:workflow} we present the workflow and define intermingledness.
The target applications for our workflow are datasets where the practitioner does not know a priori whether unique and distinguishable alternate states exist.
For demonstration and validation however, in S\ref{sec:applications} we apply the workflow to data that have already been analysed by topical experts, using three diverse climate datasets: Atlantic Meridional Overturning Circulation, Mid-latitude Coupled Ocean-Atmosphere Flow, and Habitability of Exoplanets.
Finally in \S\ref{sec:discussion} we discuss the applicability of the method in a variety of scenarios as well as existing limitations.
The workflow is also provided as easy-to-use open source Julia code that inputs NetCDF files, while the intermingledness metric has been also implemented in the global continuation functionality that tracks various measures of resilience~\cite{Datseris2023FrameworkGlobalStability, Morr2026} in the DynamicalSystems.jl Julia package.

\section{Workflow}
\label{sec:workflow}

\begin{figure}
    \centering
    \includegraphics[width=\linewidth]{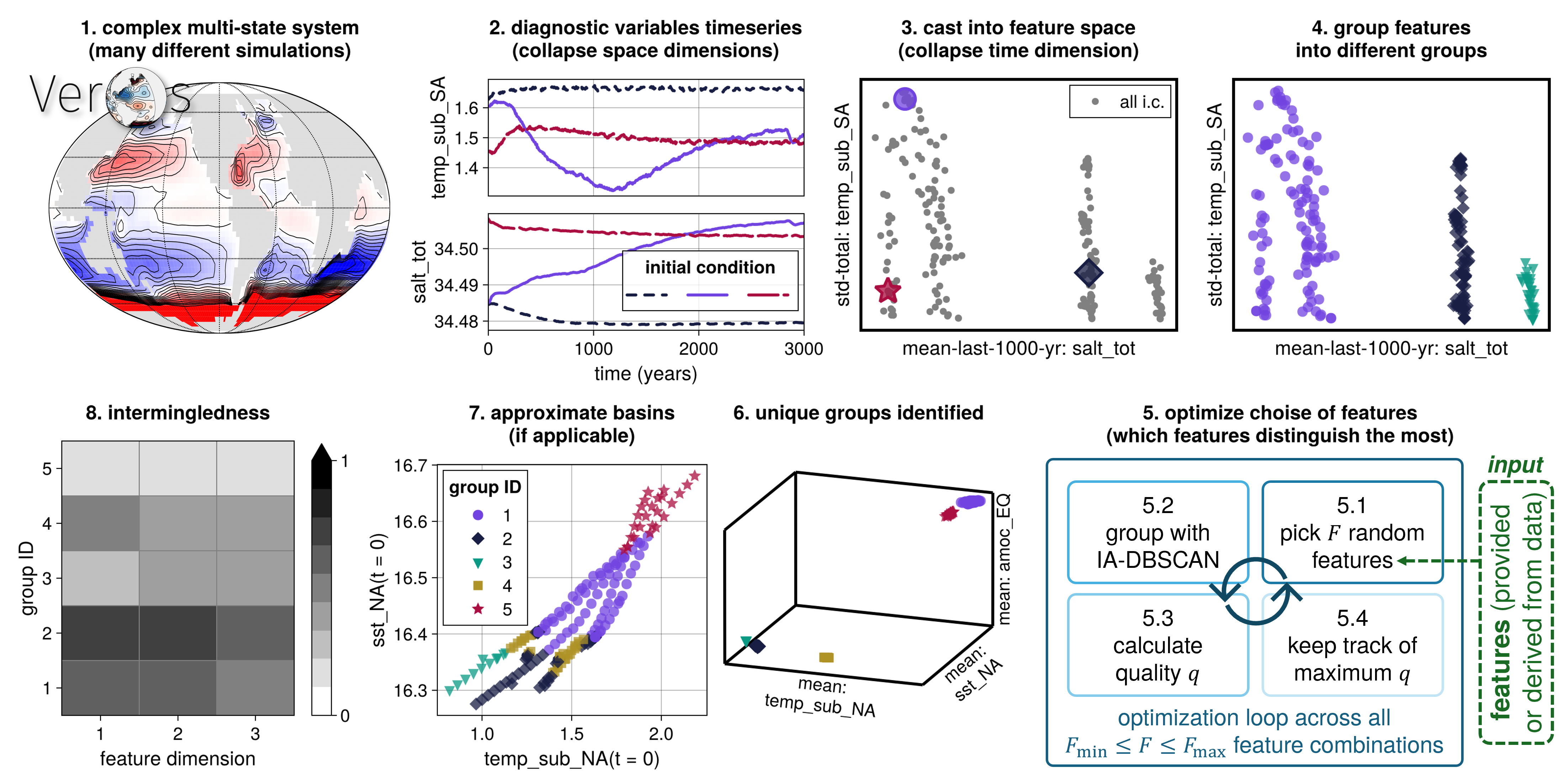}
    \caption{Summary of the steps followed by the workflow proposed in this paper, using as an example the Veros model and the dataset used in \S\ref{sec:veros}. Step 1 plots the barotropic streamfunction. The axes of steps 3, 4, 6 are plotted with bolded spines to indicate that the coordinate space is different: feature space versus the state space coordinates of steps 2 and 7. The data set includes model simulations that only converge to the vicinity of an attractor right at the end of the simulation time (e.g. purple time series in step 2). The slow convergence is primarily seen in the slowest degrees of freedom. Thus, projection onto unsuitable features (e.g. standard deviation of the entire time series of a very slow variable, see vertical in step 3) gives a poor clustering, which is then improved iteratively in step 5 yielding features that optimally separate all unique alternate states (here attractors, step 6).}
    \label{fig:summary}
\end{figure}

The workflow we propose to identify and analyse alternate states in complex data is summarized in Fig.~\ref{fig:summary} and explained in the following subsections.

\subsection{Data generation}
Step 1: The workflow starts from some complex model, often spatiotemporal. Simulations are produced from this model by altering a starting configuration of the model.
When studying multistability, this means sampling different initial conditions.
For other scenarios such as alternative operational (or response) classes, this may mean changing parameter configurations (or sets of forcings).
Regardless, this step implies a sufficient sampling of simulations, and the overall number required depends strongly on the specific model and configurations explored. We provide a short discussion about this in \S\ref{sec:discussion}.

Even when assuming sufficient simulations have been performed, the data we target are so high dimensional that traditional tools for studying multistability or alternative states will likely fail.
What follows in Step 2. therefore is a \emph{dimension reduction} step, where various \emph{diagnostic} variables are defined and extracted from the large dataset by aggregating or projecting over the space dimensions.
Each initial condition generates a set of diagnostic variables that each only depends on time. E.g., global mean temperature extracted from the spatiotemporal temperature field.
The data that remains is formally a \emph{projected dynamical system}.
Therefore, it can be directly composed with existing frameworks~\cite{Datseris2023FrameworkGlobalStability} for analysing multistability, which can also be used for other types of alternative states analyses as we show in \S\ref{sec:samosa}.
However, a key assumption we place is that the data is final: one cannot go back to the model and re-run simulations (due to cost or other reasons).
Thus need to make some adjustments on the frameworks described in Ref.~\cite{Datseris2023FrameworkGlobalStability} which we summarize over the next three subsections.

\subsection{Feature extraction}
\label{sec:features_proposed}
Our goal is to group the existing simulations into a small set of unique and distinguishable alternate states. In the analysis of multistability these are coexisting attractors of multistable systems.
Our central assumption in this process is that there has to be at least one \emph{feature} of the alternate states for which they are well distinguishable.
If such a feature does not exist, the dataset does not contain unique and distinguishable alternate states.

We formalize this in Step 3. with the concept of the \emph{feature space}. Each trajectory of the projected dynamical system is transformed into a \emph{feature vector}: a collection of statistics extracted from the timeseries that describe various aspects of the trajectory.
For example, a feature could be the global mean temperature averaged over the final 10 percent of a simulated timeseries, which likely indicates the temperature the model has stabilized at, or the overall variance of the maximum temperature, which indicates the variability in extremes.
The assumption is that if the feature dimensions are well chosen, feature vectors belonging to the same alternate state cluster together so that they form a distinct cluster (a cloud of points) in feature space. This is highlighted in Step 3. of Fig.~\ref{fig:summary}.

Naturally, the choice of features is tailored to the dataset at hand. Ideally many features should be substantially different across attractors, although the algorithmic necessity is that at least one is.
When applied to new data, and depending on context, it is likely advantageous to first detect a transient period that will be skipped~\cite{Anderson2011statistical, Pal2017practical}
This detection can be done on a per-simulation basis and the transient time itself can be considered as a feature.
From the remaining (non-transient) part of the timeseries, there are ample ways to cast it into features, such as (and each of the following could be applied to any of the diagnostic variables):
\begin{itemize}
    \item \emph{distributional}: mean, standard deviation, maximum, kurtosis, extreme quantiles, ...
    \item \emph{temporal}: main frequency or period, spectral variance or entropy~\cite{Llanos2017}, spectral weight at a predefined frequency, auto-correlation at lag 1 or 2, ...
    \item \emph{nonlinear}: sample entropy~\cite{Richman2000}, fractal dimension~\cite{Datseris2023Fractal}, recurrence quantifiers~\cite{Marwan2007}, other complexity measures~\cite{Datseris2025Complexity}, ...
\end{itemize}

In this work the extraction of relatively simple features was chosen by only considering the mean and standard deviation of the final portion of the timeseries of each diagnostic variable.
For the datasets we used this was more than enough to accurately distinguish all attractors that were already found by the topical experts.
For new applications, the practitioner should collect as many features as they think are interesting to study.
While formally the workflow runtime scales combinatorially with the number of feature dimensions $N$ (in particular $N$-choose-$F$), it is overall sufficiently fast: the whole analysis of this paper takes only $\sim$5 minutes to run on an average computer.
Thus, the total number of feature dimensions chosen does not come with big computational limitations.

\subsection{Feature grouping (clustering)}

What then needs to be done in Step 4 is to actually apply a grouping algorithm to all the generated features to decide which groups are there.
Crucially, this algorithm must not impose an a priori known number of groups, such as $k$-means~\cite{Aggarwal2018}, but estimate it on its own.
A common choice for this step is the Density-Based Spatial Clustering of Applications with Noise (DBSCAN)~\cite{Ester1996}.
It classifies points as neighbours if their distance if smaller than a radius $\varepsilon$.
Then, it clusters together points with many neighbours, and leaves as outliers points with too few neighbours.
The parameter $\varepsilon$ is crucial for the optimal operation of the algorithm. In Ref.~\cite{Datseris2023FrameworkGlobalStability} we have substantially improved the algorithm into the version we called ADBSCAN (A for Advanced) where the parameter $\varepsilon$ is optimized to yield a clustering of optimal quality.
During this process (and only for this process) all feature dimensions are also transformed into the range 0-1 based on their respective minima and maximum values, as this improves clustering performance.
For more details on this improved method, see Sect. IV. A and D of Ref.~\cite{Datseris2023FrameworkGlobalStability}.
In this work we further augmented the algorithm to IA-DBSCAN (I for Iterative), by allowing it to optionally re-apply itself recursively to each found cluster, which improved clustering for \S\ref{sec:veros}.

In Step 4 of Fig.~\ref{fig:summary} the algorithm is applied to the data plotted in Step 3. For demonstrative purposes the vertical dimension of this feature space is poorly chosen, being sensitive to transients during the equilibration time of the simulations.
As a result, it does not separate the (already known) coexisting attractors at all. But other sets of feature dimensions can be chosen, such that each unique alternate state forms a distinct cluster with a width that represents statistical fluctuations. This is achieved in the next step.

\subsection{Optimal feature selection}

Applying IA-DBSCAN directly to the whole feature space is ill advised because the feature space is likely very sparsely filled (known as ``curse of dimensionality''). In addition, we want to learn during clustering which of the feature dimensions are best suited to separate data into alternate states optimally.
In Step 5 of the workflow we, therefore, undergo an optimization loop. First, we limit the clustering to only occur in a subspace of the whole feature space composed of $F$ randomly chosen features. We allow $F$ to further run from some minimum to maximum value, and collect all combinatorial possibilities (if the number of possibilities becomes too large, a subsample of them can be selected at random).
For a specific set of $F$ randomly chosen features, we cluster the feature data using IA-DBSCAN.
This clustering is optimal with respect to the chosen features, but the choice of features itself may not be optimal.

To address this, we first record a metric $s$ that is optimized by IA-DBSCAN to yield optimal clustering. $s$ is known as the silhouette mean, see Ref.~\cite{Datseris2023FrameworkGlobalStability} for details.
Our numerical explorations highlighted that $s$ does not depend meaningfully on $F$ for small enough $F$ ($\le 4-6$), and not depend at all on the total number $A$ of attractors found for $A>1$, which we demonstrate in Figure~\ref{fig:dbscan_test}.
In the applications our workflow targets, however, we want the metric to depend on $F$ and $A$, since groupings that comprise of more alternate states and use fewer features are favourable. Thus, $s$ is augmented to define the \emph{grouping quality} $q$ as
\begin{equation}
    q = s \cdot \left[ w_A A - w_F F - w_L L \right]
    \label{eq:quality}
\end{equation}
where $L$ is the number of feature vectors that were not assigned to any cluster, but considered to be outliers, and $w_A, w_F, w_L$ are weights. In particular, as we ultimately want to uncover multi-state behaviour, $w_A$ dictates that the more unique alternate states that are found, the better, provided that they satisfy a minimum of $n_\text{min}$ points in each group.
The weight $w_F$ penalizes the number of feature dimensions based on the principle of parsimony. The fewer feature dimensions that are used to give a satisfactory separation of groups, the more informative the outcome of the algorithm is for the practitioner.
Finally, $w_L$ penalizes failing to assign a point to a cluster. A typical case where such outliers occur is if some initial conditions converge particularly slowly along certain degrees of freedom of the system. It can be useful to penalize combinations of feature dimensions that lead to a large number of outliers. At the same time, favour feature combinations that can still classify the entirety of the ensemble, despite slow convergence in some degrees of freedom.
Any of these penalty terms can be dropped by the practitioner by simply setting the corresponding weight to 0.
In Appendix~\ref{app:clustering_quality} we discuss the definition of $q$ further, providing numerical evidence that motivate it and establish its robustness versus the weights $w_A, w_F$.

\subsection{Alternative states and basins}
\label{sec:basins}
In Step 6 of Fig.~\ref{fig:summary} the optimization loop has finished and yielded a set of features that are optimal in separating the data in unique alternative states according to the metric $q$.
If studying multistability, and hence analysing timeseries of different initial conditions converging to different attractors, then the same labels can be used to recreate a sparse covering of the basins of attraction.
This is simply the value of all diagnostic variables at time 0 (or some average of the earliest steps of the simulation) colour-coded by the labels.
From there a plethora of analysis sought out by the practitioner can be performed, but here we will focus on two particular aspects relating on how well separated are the alternate states and their basins of attraction (when applicable).

\subsection{Intermingledness}
\label{sec:intermingledness}
To explore the degree of separation of alternate states and the uncertainty of their basins of attraction (when studying multistability), we define a new quantity that we call ``intermingledness''. Intermingledness describes how mixed up or shuffled (as opposed to well separated) the different groups are along the different diagnostic variables or feature dimensions.
To create intermingledness, we were inspired by the concept of high final state sensitivity~\cite{Grebogi1983} in dynamical systems theory: whether different initial conditions have high or low uncertainty about which attractor they end up at.
This concept is typically tied to whether the basins of attraction are strongly mixed with each other (fractal~\cite{McDonald1985} or ``intermingled'').
Ultimately however, our intermingledness is a fundamentally new quantity because we are in a fundamentally different context: the projection of extremely high dimensional state spaces to a few aggregating feature dimensions.

Crucially, not only the basins of attraction can be intermingled, but also the alternate states themselves (or to be precise, the features corresponding to the alternate states).
This is expected to be the case when studying alternate states beyond multistability (where the concept of basins may not even be definable), as in e.g., different parametric configurations of the system like in \S\ref{sec:samosa}.
But even in the scenario of multistability, the attractor representation in feature space can be intermingled.
This can happen because (1) feature vectors can have a (potentially large) spread across the ensemble even when belonging to the same group because of the finite time of the simulation and/or slow convergence for some feature dimensions; (2) some feature dimensions can be similar or identical between different attractors regardless of convergence, due to e.g., system symmetries.
In such scenarios it is important to know which feature dimensions are similar between alternative states, as these would be practically useless for e.g., early warning signals.

Intermingledness can provide this information, and is defined as follows. Let $U$ be a set of points, either in feature space or in basin space if applicable; $\ell$ a set of labels corresponding to each point; and $d(u, w) \in \mathbb{R}_{\ge 0}$ a distance (metric) between two points $u, w \in U$. Then
\begin{enumerate}
    \item Separate $U$ into the found groups, $U_a = \{ U[j] : \ell[j] == a, j\in \{1,...,N\}\}$ and $a \in \{1, ..., A\}$. The notation $\cdot[x]$ means the $x$-th element of the set.
    \item Let $$D(U_a, U_b) = \frac{1}{|U_a|\cdot|U_b|} \sum_{u\in U_a} \sum_{w\in U_b} d(u, w)$$ be the pairwise-averaged distance between members of group $U_a$ and members of another group $U_b$. If $a=b$ we call this distance the intra-group distance of group $a$, otherwise it is an inter-group distance.
    \item For the $a$-th group, we define as intermingledness the quantity $$I_a^d = \mathrm{summary}\{D(U_a, U_a) / D(U_a, U_b) \quad \mathrm{for}\quad b \ne a\}$$ where ``summary'' is a summarizing operator such as the mean or the maximum. In this paper we used the mean, while maximum would highlight the intermingledness between the closest groups.
\end{enumerate}
In a nutshell, intermingledness is the ratio of the pairwise-averaged intra-group distance divided by the pairwise-averaged inter-group distance.
$I_a$ is generally greater than 0 and smaller than 1, although it is difficult to assign meaning to the absolute value of $I$ without taking into account the exact shape of the basins (a plethora of additional examples of intermingledness values is provided in \S\ref{app:basin_entropy}).
If the value of $I_a$ is close to 1, it means that for the $a$-th group the corresponding points are as close (according to $d$) to those belonging to the same group as they are to those belonging to a different group.
Said differently, a point in an intermingled group ($I_a \approx 1$), when perturbed with a perturbation of size comparable to the group size, is just as likely to remain in the group than to be brought into a different group.
As such, an intermingled group cannot be distinguished easily from other groups on the basis of $d$.
Note that $I^d$ is a quantity describing the whole group; in Appendix C (\S\ref{app:basin_entropy}) we discuss a modification focused exclusively on the group's boundary, that provides some advantages and disadvantages compared to the version used in the main text.
In the same appendix we discuss more broadly potential limitations of  intermingledness, and compare it with the established concept of basin entropy.

Now the question is what distance function $d$ to use.
We decided to not utilize traditional distances (such as Euclidean) for two key reasons.
First, because the target application data are often heterotypic. As such, multi-dimensional metrics often do not make sense. E.g., how does one define a distance on the joint space of temperature and humidity?
Second, we want to be able to discuss the intermingledness of each feature dimension separately, to highlight specific aspects that are particularly (dis)similar across alternate states.
Therefore, in this paper we used many different distance functions, each corresponding to the 1-dimensional distance when projected to each of the dimensions in the set $U$. As a result, intermingledness is defined individually for each feature dimension.

Throughout this paper intermingledness is presented in a matrix form, with the rows corresponding to the found groups, and the columns to the quantities we project the data into: either the feature dimensions, or the diagnostics when studying basins of attraction.
In Step 8 of Figure.~\ref{fig:summary} we present as an example the intermingledness of the basins of attraction of the the diagnostic variables used to produce the features shown in panel 6.
We can see, for example, that the basin of attractor 2 is particularly intermingled: when projected to any of the three diagnostic variables, the initial condition for that variable is as close to belonging to any other basin than its own basin (also visible in panel 7).

In the applications that follow, we apply intermingledness to both the alternate state features, and the basins of attraction when applicable.
This allows us to learn (1) which feature dimensions are most or least similar between alternate states and (2) which diagnostics best discriminate to which attractor initial conditions converge to (those with small intermingledness).

\subsection{Continuation}
\label{sec:continuation}
Continuation is the process where one tracks the model attractors of a multistable model, as a parameter is varied.
This is straightforward to do within the methodology we propose here, by utilizing a \emph{matching algorithm} proposed in Ref.~\cite{Datseris2023FrameworkGlobalStability}. Here is the summary:
First, the same multistability analysis (steps 1-6) is repeated for every different parameter(s) the simulations have been performed at.
Then, the attractors (groups) found at different parameters are ``matched'' with each other according to a chosen measure of similarity.
Matching means that attractors that are similar enough, but exist at different parameters, get labelled (i.e., colour-coded) with the same integer.
In this work, the chosen measure of similarity is the distance of the group centroids when projected to the dimensions of one of the best features.

The result of this matching process can then be applied further to any other quantity that has been computed across the parameter axis.
In the application of \S\ref{sec:atmos} we showcase this by tracking not only the attractors but also their intermingledness (similarity of the features representing each attractor).

\section{Applications}
\label{sec:applications}

Here we present the application of the workflow to three different datasets.
The first two are targetting multistability analysis, where different initial conditions are varied.
The third highlights the applicability of intermingledness in alternate operating states originating through varied parameters.
For the second dataset we highlight how the workflow can be applied over a continuation.
Each subsection also summarises briefly the simulation setup of the data, while more details are provided in \S\ref{app:model_details}.

\subsection{Case 1: Atlantic meridional overturning circulation}
\label{sec:veros}

Simulations with the ocean model Veros show a multistability of the Atlantic meridional overturning circulation (AMOC), where steady states with a vigorous circulation can co-exist with states of fully or partially collapsed circulation, in agreement with previous studies \cite{Rahmstorf1995}. As presented in previous publications \cite{Lohmann2024AMOCMultistability, LOH24b}, these large ensemble simulations reveal a hitherto unknown degree of multistability, with coexisting steady states that differ only in subtle ways, including states with similar circulation strengths of the AMOC but contrasting spatial patterns or spatio-temporal variability. Some of the coexisting states are a result of alternative spatial patterns of convection \cite{Lohmann2024AMOCMultistability, ThiesDijkstra2026}. This high multistability challenges the paradigm of a potential future AMOC collapse at a singular tipping point, and its predictability by monitoring changes of the local stability of a single present-day attractor.

Obtaining this extensive dataset and deducing the multistability involved substantial manual effort, which could be alleviated and refined by the automated approach presented here. The data was obtained firstly by manual continuation of initial conditions along different branches of attractors, and secondly by discovering new stable steady states using ensemble simulations that interpolate between previously known branches (see Appendix~\ref{app:Veros_details} for more details). The N=259 equilibrated simulations were classified by hand according to their asymptotic state, with additional cross-checking by extra-long control simulations (20,000+ years). Since it is a high-dimensional system, substantial effort was needed to check the differences of the asymptotic ensemble states in as many observables as possible.

\begin{figure}
    \centering
    \includegraphics[width=\linewidth]{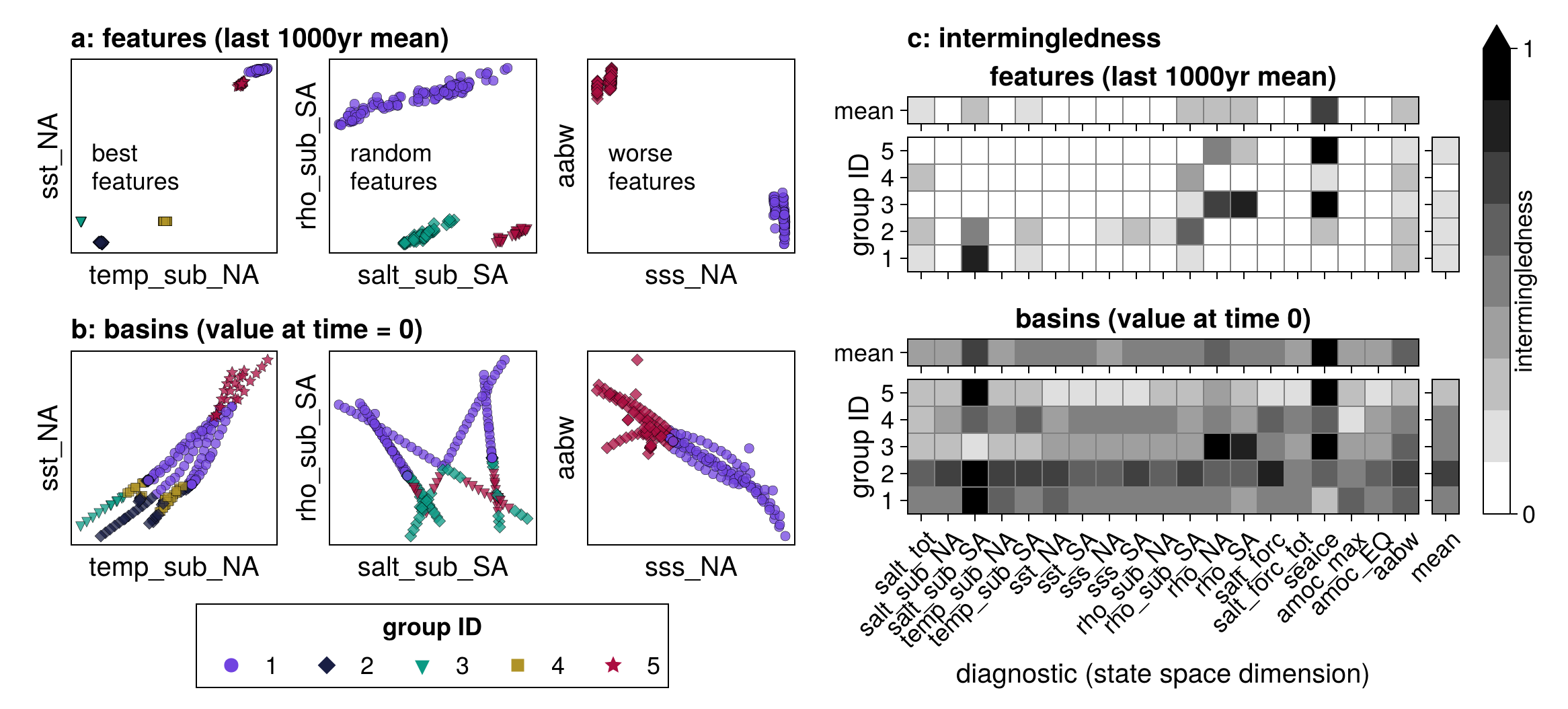}
    \caption{Analysis of Veros data. The abbreviations of the diagnostic variables ('sst\_NA', 'temp\_sub\_NA', etc.) are explained in Appendix~\ref{app:Veros_details}. \textbf{a}: Scatterplot of features, which are the mean of the last 1000yr of simulation, for various diagnostic variables. The first panel depicts the optimally chosen features that best separate the data into attractors.
    The second and third panels show randomly chosen and worst performing features (on the basis of $q$), respectively. The scatterplots are colour-coded by the clustering resulting from choosing the particular features. \textbf{b}: Same, but for the basins of attraction (i.e., value of the diagnostics at time = 0). \textbf{c}: intermingledness for both the features of the diagnostic variables (i.e., intermingledness of the attractor features) and for the basins (i.e., intermingledness of the initial conditions).
    Paradigmatic timeseries for this dataset is shown in Step 2 of Fig.~\ref{fig:summary}.}
    \label{fig:veros}
\end{figure}

The automated approach is able to recover the grouping into five distinct attractors and finds relevant, non-obvious observables that can distinguish the different states (Fig.~\ref{fig:veros}a).
While the AMOC strength itself is viable at distinguishing the attractors (see low intermingledness in top of Fig.~\ref{fig:veros}c), the best features AMOC strength combined with North Atlantic surface and sub-surface temperature (Fig.~\ref{fig:veros}a) (note: only two of the three are plotted in the figure).
This is in line with the finding that at least some of the coexisting states are a result of different feasible levels of North Atlantic convective activity \cite{Lohmann2024AMOCMultistability, ThiesDijkstra2026}, whereby the surface is more or less coupled to the deep ocean.

The basins of attraction have intermingledness substantially less than 1 in most diagnostics (Fig.~\ref{fig:veros}c bottom), which is also evident from panels (b), noting that intermingledness is calculated along one dimensional projects.
This reflects that fact that while none of the diagnostic variables can perfectly distinguish the different groups on their own, the basins are not completely mixed either.
When averaged across features, the basins corresponding to vigorous AMOC states (group ID 1 and 5) show clearly lower intermingledness compared to those with collapsed AMOC.
This could indicate a higher predictability in the vigorous versus the collapsed AMOC regime.
Intermingledness is also smaller when calculated in a two dimensional space (not shown), which highlights that multiple diagnostics separate basins better than their one-dimensional projections (something expected as the workflow finds 3 diagnostics as best distinguishing the alternate states).


Consistently high values of intermingledness across diagnostics and groups can be a good indicator of fractal basins of attraction (here potentially for groups 2-4). Although for the present model setup it has of yet not been shown \cite{LOH24b}, a slightly different model configuration was shown to possess a fractal basin boundary \cite{Lohmann2021}. We refer to \S\ref{app:basin_entropy} and remind that the absolute value of intermingledness on its own is difficult to interpret, only its closeness to 1 is.

\subsection{Case 2: Mid-latitude Coupled Ocean-Atmosphere Flow}
\label{sec:atmos}
The MAOOAM (Modular Arbitrary Order Ocean Atmosphere Model)~\cite{decruz2016a, demaeyer2025a}, with a non-linear longwave radiation scheme, produces multiple stable attractors. These correspond to multiple distinct long term repeating flows in the coupled ocean-atmosphere system, commonly referred to as Low Frequency Variability (LFV)~\cite{Hamilton2023}. These distinct attractors coexist for different levels of coupling between the low order quasi-geostrophic atmosphere and the single layer ocean, as well as for different levels of atmospheric emissivity ($\varepsilon$), which is the proxy for climate change in this simple model. In Ref.~\cite{Hamilton2023} three distinct attractors were found, which coexist for particular parameter values. These distinct attractors were classified by calculating the mean ocean and atmosphere temperatures, as well as visually inspecting the LFV over particular modes of the model. These modes were selected as the model proxies of the real world North Atlantic Oscillation, as they present a similar north-south double gyre oscillation~\cite{vannitsem2015}.

In this study we produced trajectories from differing initial conditions, for different values of atmospheric emissivity ($\varepsilon$), to compare the results of the methods outlined in this paper, against those previously published.
Further details of the model and the model integrations can be found in \S\ref{apx:maooam}. For this study the last 25\% of each trajectory was used, and the feature chosen was the standard deviation, which produced better results than the mean.
Figure~\ref{fig:atmos}-a. shows that the method identified three distinct attractors, matching published results. Interestingly, the method found that different modes ($\psi_{\mathrm{o}, 6}$ and $\psi_{\mathrm{a}, 3}$, corresponding to particular ocean and atmosphere streamfunctions, see \S\ref{apx:maooam} for further details) produced optimal clusters, compared with those used in the original paper. Figure~\ref{fig:atmos}-b. shows that there is no clear separation in the basins, given particular modes of the model, a result backed up by the high intermingledness of the basins, shown in Figure~\ref{fig:atmos}-c.
This is a classic result from chaotic systems, presenting fractal basins of attraction. Lastly, we see that particular features, namely $\psi_{\mathrm{a},1}$ and $\theta_{\mathrm{a}, 1}$, have an intermingledness close to 1, again indicating fractal basins.
This is likely a result of the atmosphere modes having a similar behaviour over these variables, which are driven by the difference in thermal forcing between the south and north of the domain.

\begin{figure}
    \centering
    \includegraphics[width=\linewidth]{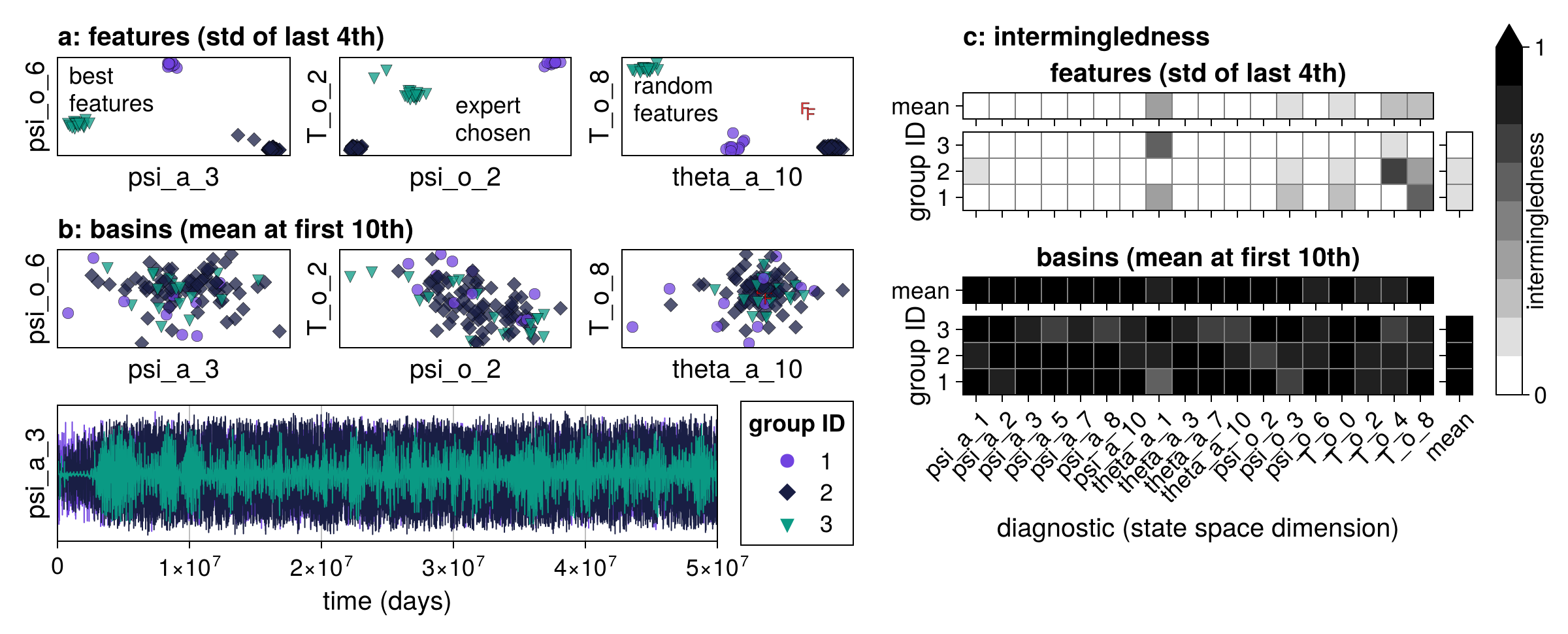}
    \caption{As in Fig.~\ref{fig:veros}, applied to the data produced by the MAOOAM model. The feature clusters (a) show that while there are three distinct clusters, the best features to cluster over were different to those used in the original paper. Panel (b) shows the fractal basin boundaries of the attractors, from the initial conditions. An additional panel at the bottom left of the figure showcases time-series for an observable of this simulation set, showing distinct low frequency variability, or quasi periodic behaviour over long time periods, between the different groups or attractors. In the model this results in different long term climatic patters with different principal periods. Panel (c) shows the intermingledness of the features and basins. The basins show a high level of intermingledness - which is unsurprising given the fractal nature of the basins as shown in panel (b).}
    \label{fig:atmos}
\end{figure}

In Fig.~\ref{fig:atmos_cont} (top panel) we present the continuation for this dataset. The continuation method presented in this study can reproduce the result found in~\cite{Hamilton2023} where one of the three attractors (shown in green) appear to destabilise for $\varepsilon<0.85$. In Fig.~\ref{fig:atmos_cont} (bottom panel) we present how the mean intermingledness measure changes as part of the same continuation. As shown, the mean intermingledness for group 3 (green squares) increases considerably past $\varepsilon > 0.9$, crossing the other two attractors. This means that for increasing emissivity we see that one of the three attractors becomes less distinguishable from the others. This could point to an approaching bifurcation, where the green attractor collapses into one of the other existing attractors. As the value of $\varepsilon$ goes to $1$ the atmosphere radiates heat as a perfect black body, leading to radiation becoming more dominant in the heat transfer, which impacts the dynamics of the coupled system.

\begin{figure}
    \centering
    \includegraphics[width=0.5\linewidth]{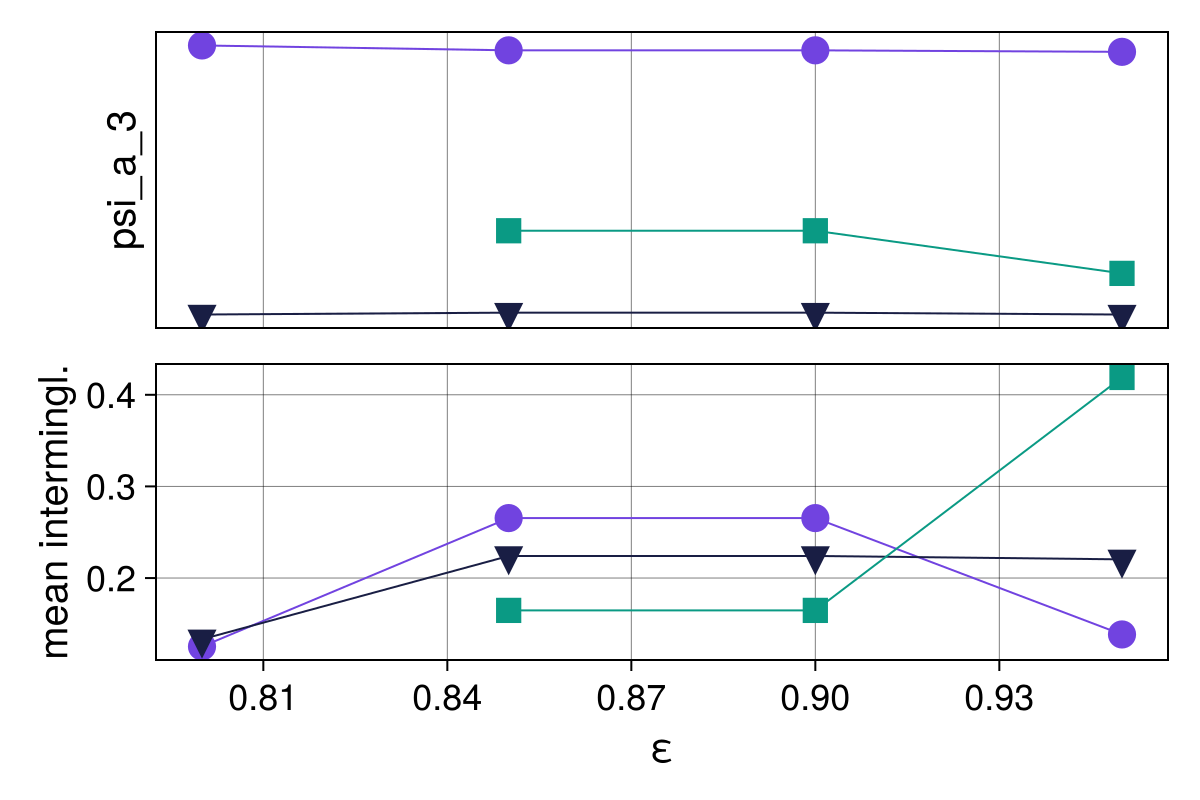}
    \caption{Continuation, as defined in \S\ref{sec:continuation}, for the MOAOOAM dataset. Top panel: the centroids for each feature group (attractor) for varying the atmospheric emissivity $\varepsilon$, projected to the diagnostic variable \texttt{psi\_a\_3}. This atmospheric mode corresponds to the barotropic streamfunction projected on an east-west gyre mode. The result is that larger or smaller centroids on this variable correspond to the a larger or weaker average strength of the west-east aligned pressure systems in the atmosphere. Bottom panel: intermingledness of features, for varying $\varepsilon$, averaged across all diagnostics.}
    \label{fig:atmos_cont}
\end{figure}

\subsection{Case 3: Habitability of Exoplanets}
\label{sec:samosa}

The final dataset we explore is used in SAMOSA (Sparse Atmospheric Model Sampling Analysis), which is an exoplanet model intercomparison project focused on a scenario of a hypothetical Earth-sized planet in synchronous rotation around a low mass star~\cite{HaqqMisra2022}. This configuration leads to a substellar hemisphere of the planet in perpetual daylight and an anti-stellar hemisphere in perpetual night, which results in unique and non-Earthlike circulation patterns and nonlinear responses to changes in stellar or atmospheric forcing. Studying such model configurations is of interest to support the planning and interpretation of exoplanet observations with space- and ground-based telescopes, with low-mass stars providing the most numerous and accessible targets. The dataset presented here was generated using the ExoPlaSim general circulation model of intermediate complexity across a 2-axis parameter space of incident stellar radiation (instellation) and total surface pressure~\cite{paradise2022exoplasim}. Each of the simulations involves a unique combination of these two parameters, with all other parameters identical at initialization, and the model then integrated until reaching a statistically steady state (see \S\ref{sec:exoplasim} for details). The SAMOSA intercomparison is intended to evaluate models of greater (and lesser) complexity against the results obtained from this intermediate-complexity model.

\begin{figure}
    \centering
    \includegraphics[width=\linewidth]{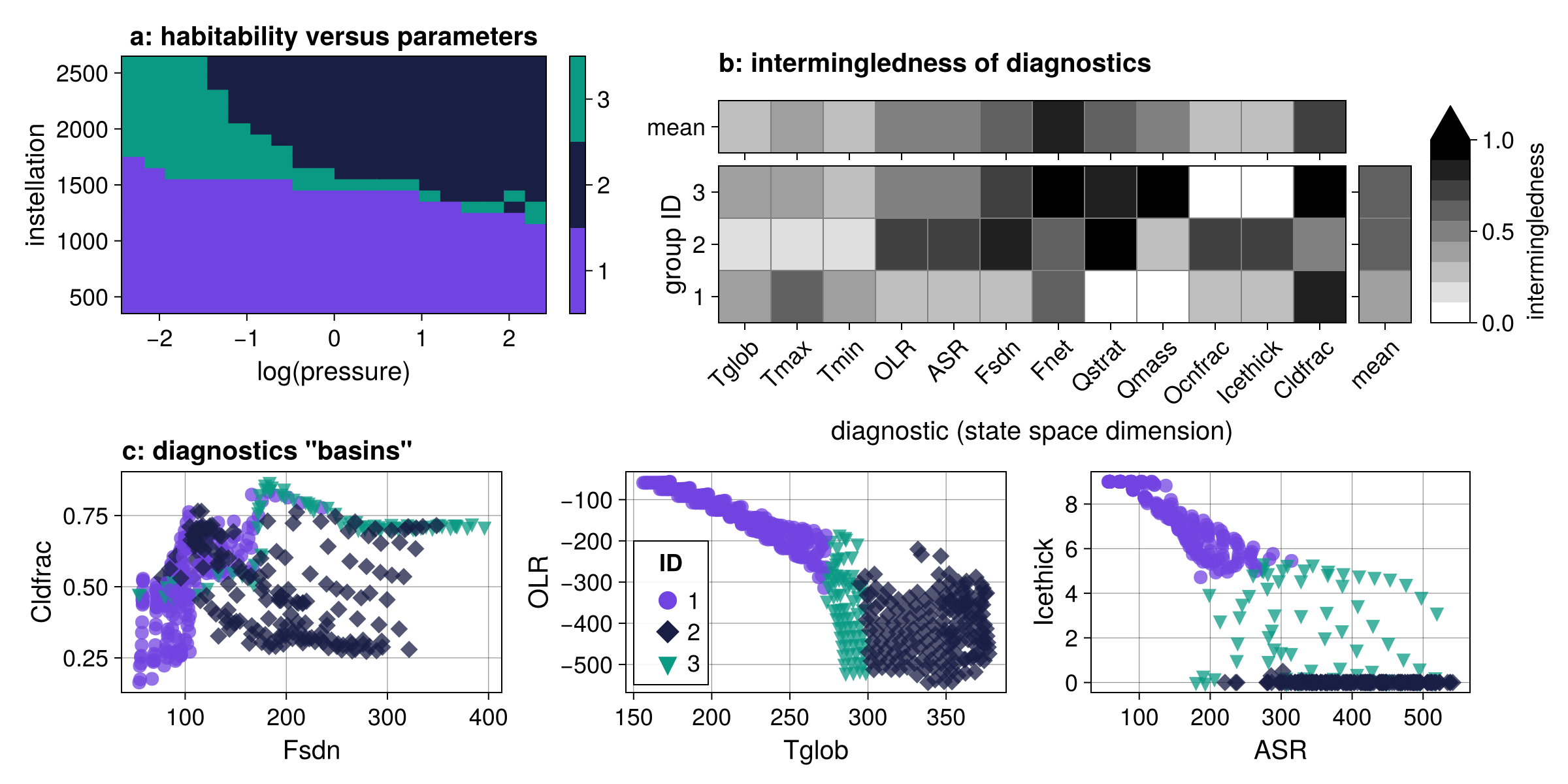}
    \caption{Analysis of data produced by the ExoPlaSim model for the habitability of an exoplanet around a low-mass star, used in the SAMOSA intercomparison. \textbf{a}: Three colour-coded habitability classes are found across the parameter space of instellation and pressure. \textbf{b}: Intermingledness of the of the features (global and time mean of each diagnostic variable). \textbf{c}: A random selection of features colour-coded by the habitability classes.}
    \label{fig:samosa}
\end{figure}

This dataset is not a multistability study like the previous two, but instead a study of different parameter configurations.
Alternative states in the data exist nevertheless, and are already classified: each simulation is assigned to a particular habitability type.
The habitability types are defined as binary flags based on the temperature (global mean or minimum) exceeding the freezing point of water (273.16\,K) as well as on the temperature (global mean or minimum) remaining below general thermal limits for biology (323\,K), which partition into (i) icy, (ii) warm, and (iii) hot climate regimes (Fig.~\ref{fig:samosa}a) that are consistent with prior analysis~\cite{paradise2022exoplasim,HaqqMisra2022}.
As such, there is no need for Steps 4-6 of our methodology.
This is not to say that steps 4-6 are not possible; one could define features that capture aspects affecting habitability such as temperature and water. It is just that this processing has already occurred for this dataset, and the data provided were already temporal means of all diagnostic variables (i.e., the features were provided directly instead of being extracted from timeseries).

What we can do instead is directly apply the concepts of intermingledness to the data, and discuss what we learn from this.
The intermingledness of diagnostics (Fig.~\ref{fig:samosa}b) shows some diagnostics with several high values at or near 1, some low values at or near 0, and many intermediate values. It is interesting to note that some diagnostics show very low intermingledness for one group but very high intermingledness in another group. The basins in Figure~\ref{fig:samosa}c show the groups falling into three well-demarcated basins for diagnostics with relatively low ($\lesssim 0.5$) mean intermingledness (middle and right), whereas the diagnostics with higher mean intermingledness do not show clear separation of the basins. In the specific case shown (Figure~\ref{fig:samosa}c, left), the high-intermingledness cloud fraction (Cldfrac) diagnostic variable is an average quantity across all vertical cloud decks, which may not have a direct relationship with any of the habitability parameters. This suggests possible applications for intermingledness in model intercomparison studies across parameter spaces, such as identifying the most useful diagnostics based on basin separation and then focusing on these diagnostics for further analysis and comparison. Comparing the intermingledness of different models across the same parameter space could also reveal important differences or similarities in the model's results.

\section{Discussion}
\label{sec:discussion}

\subsection{Further application scenarios}
\label{sec:further_scenarios}

Beyond what we have highlighted so far, we can imagine several additional applications of the workflow. The information on important observables for distinguishing alternate states can help guide the planning of observational campaigns where only a limited number of observables can be measured.
The same information may also help define indicators for critical transitions between the alternative states~\cite{Scheffer2009}.
On the one hand, for tipping points caused by a bifurcation or crisis, statistical early-warning signals  are expected to be most pronounced in diagnostic variables that are somewhat orthogonal to the basin boundary and that are good at separating the attractors \cite{Lohmann2025EdgeStatesEWS, LOH25b}. This would be diagnostic variables or features with low intermingledness in basins and attractors.
On the other hand, when considering noise-induced transitions in the multistable regime away from a bifurcation, diagnostic variables with high intermingledness in the basin of attraction may indicate directions in which a perturbation is more likely to cross a basin boundary leading to a critical transitions. Note that this requires the intermingledness quantity to reflect a truly intermingled or tightly folded nature of the basins. This does not have to be the case, for instance when using distance functions that are restricted to only one or a few feature dimensions or diagnostic variables (e.g. the horizontal dimensions Fig.~\ref{fig:veros}b could make the purple basin appear intermingled).

Another usage case is model intercomparison studies that care about multistability.
Further habitability intercomparison projects~\cite{sohl2024cuisines} for example may benefit, as can tipping-oriented intercomparison projects like TipMIP~\cite{TipMip}.
Besides the analysis of each entry in the intercomparison projects, this workflow can also establish objectively whether the models in the intercomparison have similar attractors to begin with (in terms of their features) and that they perform similar transitions to other attractors (the continuation matching procedure of \S\ref{sec:continuation} can also be applied across datasets).
This can be done by matching (\S\ref{sec:continuation}) the attractors across the entries of the intercomparison instead of across a parameter.

Another application scenario is what has been called ``stochastic bifurcations''~\cite{Romanou2023}, or ``ensemble splitting'~\cite{Boerner2025}, which has been observed in a variety of complex models~\cite{Gu2024-cd, Mehling2024AMOCBoxLorenz, Lohmann2024Storylines, Lohmann2021, Kaszas2019-ub, Oh2025}.
Even if the exact mechanism may differ, all these cases are examples of final state sensitivity: the fact that even minute changes in the initial configuration or forcing trajectory of the model may lead to completely different end states.
Intermingledness allows one to better understand which changes precisely may alter the final state (i.e., for which diagnostics the basins are most intermingled).

Applications beyond multistability are also possible and useful. For example, consider the study of power grids which are known to display a plethora of alternate states~\cite{Hellmann2020}. Even when fixing the system state to the synchronous solution (desired operation of the grid), one can then study response to a large variety of perturbations, each targetting different nodes or perturbing globally different grid variables or changing the couplings of the network.
The system is expected to respond in one of few unique alternative ways in such perturbations such as, black out, partial black out, return to desired operation, or taking too long to stabilize back to a steady state.
Our workflow is directly applicable in such a scenario to distinguish first what are the unique alternative responses and then what aspects of perturbations separate the system response best, which would likely lead to more focused analysis.

Lastly, another key application is what was also touched upon in \S\ref{sec:samosa}: that intermingledness can highlight model deficiencies (often also called biases).
If theoretical knowledge dictates that a particular diagnostic variable \emph{should} affect the model final state, but at least according to intermingledness it does not, this is a good indicator that some model bias or insufficiently accurate parameterization may be at play.

\subsection{Limitations}
\label{sec:limitations}
The main limitations of the presented workflow are: number of simulations necessary; difficulty with long transients; interpreting intermingledness.

The workflow requires multiple simulations because this is required for clustering with IA-DBSCAN.
The higher the number, the more accurate the classification into alternative steady states, but on the other hand one would like the minimum optimal number, as the applications we target are dominated by data sparsity due to cost.
Unfortunately, one cannot know a-prior how many will be ``good enough'', as this will be hugely varying from system to system and from context to context (i.e., multistability vs. parameter variability).
Beyond the system sensitivity, this downside is also a limitation of DBSCAN being ``black-box-like'' in the sense of not being able to judge why it won't perform well a-priori (this is a limitation that is discussed further in Ref.~\cite{Datseris2023FrameworkGlobalStability}).
For the data we collected for this paper, we sought out a number of simulations on the order of 50-100, which ended up being more than sufficient.
The benefit of having more simulations is that a statistically meaningful estimation of intermingledness can be done and, when applicable, the basins can be explored as well.
For Earth System sciences, arguably models of intermediate complexity~\cite{Claussen2002, Weber2010} are best suited for the proposed workflow. For other scientific disciplines it could be computationally more achievable to generate ample initial conditions (for example power grid models have typically 1000s degrees of freedom and showcase multistability of synchronous and asynchronous states).
Note that one could also process datasets that consist of simulations at many different parameter values, but only few initial conditions each. The outcome of the workflow would then indicate whether critical transitions occurred as a result of the parameter variation, and which observables are most impacted.

Regarding transients, it is possible that the workflow may classify transients as steady states for particular choice of features.
This scenario could be desired or undesired by the practitioner. E.g., in Ref.~\cite{Boerner2025} identifying transient states was a focal point, however in climate sensitivity type of studies we imagine it is undesirable to classify transients as steady states.
On the other hand, this can be undesirable if the transients are relatively short, and one ends up artificially introducing ``duplicate'' attractors because some initial conditions take a bit longer to converge to the same attractor.
Thus, often care needs to be taken to identify a transient period in the data.

The intermingledness indicator has its limitation when interpreted quantitatively, especially for low values. When applied to the basins of attraction, its value for a given basin $a$ depends on the relative size (or extent of sampling) of basin $a$ compared to the other basins. Thus it could be for instance that basin $a$ and basin $b$ are equally well separated from all other basins along a particular feature dimension (i.e. not intermingled), but that basin $a$ is smaller (or sampled less extensively) than basin $b$ and thus yields a smaller value of the intermingledness indicator.
When interpreting the intermingledness of basins one generally has to be aware that high values do not necessarily represent a true intermingledness of the basins, if the underlying distance function $d$ was computed on only a sub-set of diagnostic variables, or a basin of attraction is so small that it is impossible for it to be intermingled.

\section{Conclusions}
In this paper we proposed an objective workflow for distinguishing alternate states in a potentially multi-state complex dataset.
At its core, it is applicable for any simulation set which may contain alternative states, even if not traditionally multistable, as was discussed in \S\ref{sec:further_scenarios}.
The workflow also provides a simple-to-interpret quantifier for highlighting across which diagnostic variables (or more precisely feature dimensions) the alternate states are most or least similar, and across the initial conditions (or generally different configurations) of which diagnostic variables the basins are most or least intermingled.
An immediate benefit of that is additional guidance for the researcher(s) or broader community: often analysis may be biased in a couple of observables (such as temperature in the case of climate), however these need not be the observables that best distinguish alternative states, or the observables whose perturbations are the most likely to cause a transition between alternative states.
We believe our workflow will accelerate, attribute robustness, and enable new findings, for a plethora of analyses that relate to unique alternate states of a complex and high dimensional system.

\section{Data and code availability statement}

The raw simulation data of the Veros model are deposited in the Electronic Research Data Archive repository of the University of Copenhagen at \url{https://sid.erda.dk/sharelink/BMov9nurth}.
The MAOOAM model~\cite{demaeyer2025a} is open source and can be accessed at \url{https://github.com/Climdyn/qgs/tree/v1.0.0}.
The dataset produced using the MAOOAM model is not open access but available upon request.
The exoplanet habitability dataset is available in \cite{ExoPlaSim}.

The code that reproduces the figures of this article is open source and also includes a guiding Jupyter notebook for applying the workflow to new data. It is available on GitHub under the URL \href{https://github.com/Datseris/ComplexMultistability}{\texttt{github.com/Datseris/ComplexMultistability}} and archived on Zenodo~\cite{codebase}.
The measure of intermingledess is available in the aforementioned codebase but it is also implemented in the Attractors.jl subpackage of DynamicalSystems.jl~\cite{Datseris2018DynamicalSystemsjl}.

\section{Acknowledgments}
We thank Alexandre Wagemakers for helpful comments on the manuscript in particular regarding intermingledness.

\section{Appendix A: model details}
\label{app:model_details}
\subsection{Veros}
\label{app:Veros_details}

The Veros ensemble at a fixed value of the control parameter F (north Atlantic freshwater forcing) is constructed by choosing equally spaced initial conditions that lie on straight lines across phase space. Each straight line is defined by a pair of states, one taken from a vigorous AMOC attractor, and the other from an attractor with collapsed AMOC. This is done for different pairs of attractors for a given F, since both the vigorous and collapsed regimes consist of multiple branches of attractors. The states A and B on the attractors that define the straight lines are chosen as the model snapshot at the last time step of the long equilibrium simulations that were performed at fixed forcing values \cite{Lohmann2024AMOCMultistability}, and thus correspond to arbitrary samples from the attractor.

For each pair of states, we perform a linear interpolation of the entire set of model state variables and
construct 32 equally spaced initial conditions C:
\begin{equation}
C = \alpha A + (1 - \alpha) B, \,\,\,\, \text{where} \,\, \alpha = (i-5)/20,  \,\,\,\, \text{and} \,\, i = {0,1,..,31}.
\end{equation}

With this choice of $\alpha$, also states on the line reaching beyond the collapsed and vigorous states are considered. Thus, a relatively large variety of initial ocean states is sampled. In the case presented here (F= 8.1), there are 2 branches of vigorous attractors, and 3 branches of collapsed attractors. As a result there are 6 pairings of a vigorous and a collapsed attractor, defining initial conditions on 6 straight lines. A selection of additional simulations where the initial conditions were obtained by other means are included in the ensemble. These include the simulations from the continuation experiments by \cite{Lohmann2024AMOCMultistability}, which were used to construct the stability landscape of the Veros model. Further, for some pairs of attractors, simulations with 10-15 initial conditions were performed that more closely sample the area around a basin boundary by repeated bisections along the same straight line \cite{LOH24b}. Taken together, this yields an ensemble with N=259 members.

As in another previous study \cite{Lohmann2025EdgeStatesEWS}, the discrete spatio-temporal fields of the ocean state (almost 1 million dimensions) is compressed into a set of observables by averaging the physical variables in different spatial boxes, and by considering the maxima of streamfunctions. For the analysis presented here we only retain observables pertaining to the Atlantic and to the thermohaline circulation (disregarding the wind-driven circulation). In the following the observables and their abbreviations used in the main text are described.
The set of observables comprises salinity, temperature and density in the north (abbreviated by 'xxx\_NA')and south Atlantic ('xxx\_SA') averaged for the surface and subsurface ocean, respectively. Surface temperatures, salinities and densities are denoted by 'sst\_xxx', 'sss\_xxx', and 'rho\_xA', respectively. Subsurface averages are 'salt\_sub\_xxx', 'rho\_sub\_xxx', etc. In addition, global average salinity ('salt\_tot'), north Atlantic and globally averaged surface salinity forcing ('salt\_forc' and 'salt\_forc\_tot'), as well as global sea ice extent ('seaice') is used. Finally, the maximum of the AMOC streamfunction below 500 meters ('amoc\_max'), the value of the AMOC streamfunction at the equator at 1000m depth ('amoc\_EQ'), and the minimum of the Antarctic bottom water cell ('aabw\, again from the meridional streamfunction) is used. This yields a set of 19 observables.

\subsection{MAOOAM}\label{apx:maooam}
The MAOOAM model consists of a simple two layer quasi-geostrophic atmosphere coupled to a single quasi-geostrophic ocean layer~\cite{decruz2016a}. The model equations are projected onto a set of basis modes, known as a Galerkin-expansion, to create a set of 38 ordinary differential equations. The modes are picked to respect the boundary conditions of the model, and the number of modes are chosen to respect the key scales of the mid latitude atmosphere. The key variables discussed in this paper are the barotropic streamfunctions of the atmosphere ($\psi_{\mathrm{a}, i}$) and ocean ($\psi_{\mathrm{o}, i}$), the baroclinic streamfunction of the atmospheric (from which we can calculate the atmospheric temperature) ($\theta_{\mathrm{a}, i}$), and the temperature of the ocean ($T_{\mathrm{o}, i}$). The subscript points to which mode the variable has been projected on. The behaviour of variables on particular modes highlight particular ocean or atmospheric flows which can then be used to understand climate characteristics of the model~\cite{vannitsem2015, hamilton2025c}.

In previous work the key variables analysed were the ocean north-south double gyres for flow and temperature ($\psi_{\mathrm{o}, 2}$ and $T_{\mathrm{o}, 2}$), as these showed similar patterns and timescales as the North Atlantic Oscillation, and are some of the clearest examples of low frequency variability in this model.

The initial conditions were chosen by taking the maximum and minimum bounds on each variable for each of the attractors found in the previous studies, and selecting points in this 38 dimensional space using a uniform distribution. The initial transient part of the trajectory was removed. The resulting model behaviour for differing values of emissivity was achieved by selecting initial conditions as above, and fixing the atmospheric emissivity before running the trajectory.

\subsection{ExoPlaSim}\label{sec:exoplasim}
The ExoPlaSim general circulation model of intermediate complexity is a modified version of the Planet Simulator (PlaSim) model~\cite{paradise2021climate} that has been extended for application to exoplanet atmospheres~\cite{paradise2022exoplasim}. The model uses a spectral dynamical core, with a 50\,m mixed-layer slab ocean and no topography for these simulations. Radiative transfer is simplified with only two shortwave bands and one longwave band, which accounts for much of the reduced computational time compared with other computationally-intensive general circulation models. The set of ExoPlaSim calculations used in this study were previously published and analyzed~\cite[][]{paradise2022exoplasim}, which consisted of 460 simulations that span a regularly-spaced grid of instellation from 400--2600\,W\,m$^{-2}$ (spaced linearly) and surface pressure from 0.1--10\,bars (spaced logarithmically). Changes in surface pressure are represented primarily as N$_2$ gas pressure, with CO$_2$ partial pressure fixed at 400\,$\mu$bar and H$_2$O abundances calculated from surface evaporation. Simulations assumed a fixed 15-day rotation period and a host star with a 3000\,K blackbody spectrum. Simulations were continued until changes in mean radiative fluxes at the surface and top-of-atmosphere remain less than 0.5\,W\,m$^{-2}$ per decade. The model treats water vapor as a minor atmospheric constituent, which is known to limit the accuracy of predictions for hot climates. Such model simplifications may be more or less significant at different regions of the parameter space; identifying such differences between ExoPlaSim and other models is one of the goals of the SAMOSA intercomparison project~\cite{HaqqMisra2022}.

\section{Appendix B: details on clustering quality}
\label{app:clustering_quality}
In this appendix we give more details on the definition of the clustering quality, Eq.~\eqref{eq:quality}.
The definition of Eq.~\eqref{eq:quality} is arbitrary in the sense of not having any theoretical foundation, and was derived here empirically.
We have not found much literature on the behaviour of optimizing DBSCAN as the number of found clusters increases.
There could be alternative definitions that have an appropriate theoretical backing and should replace Eq.~\eqref{eq:quality}.
Thankfully, the remainder of the workflow, along with all potential applications, remains unaffected by such a substitution.

In Figure \ref{fig:dbscan_test} we show the optimal silhouette mean $s$ for various combinations of $A, F$. $s$ is optimized by the ADBSCAN algorithm: a range of possible radii values $\varepsilon$ for DBSCAN is scanned, and the choice yielding maximal $s$ is kept and utilized for the final DBSCAN clustering.
In Figure~\ref{fig:dbscan_test} there is no discernible difference of the upper quantiles of $s$ irrespectively of $A$ or $F$. This result is not unique to the data used in the figure. It means that as long as the DBSCAN clustering output is optimal, the silhouette mean is optimized to a value close to 1 irrespectively of $A, F$.
This is precisely what motivated us to define $q$ and involve the weights $w_A, w_F$.

\begin{figure}
    \centering
    \includegraphics[width=0.5\linewidth]{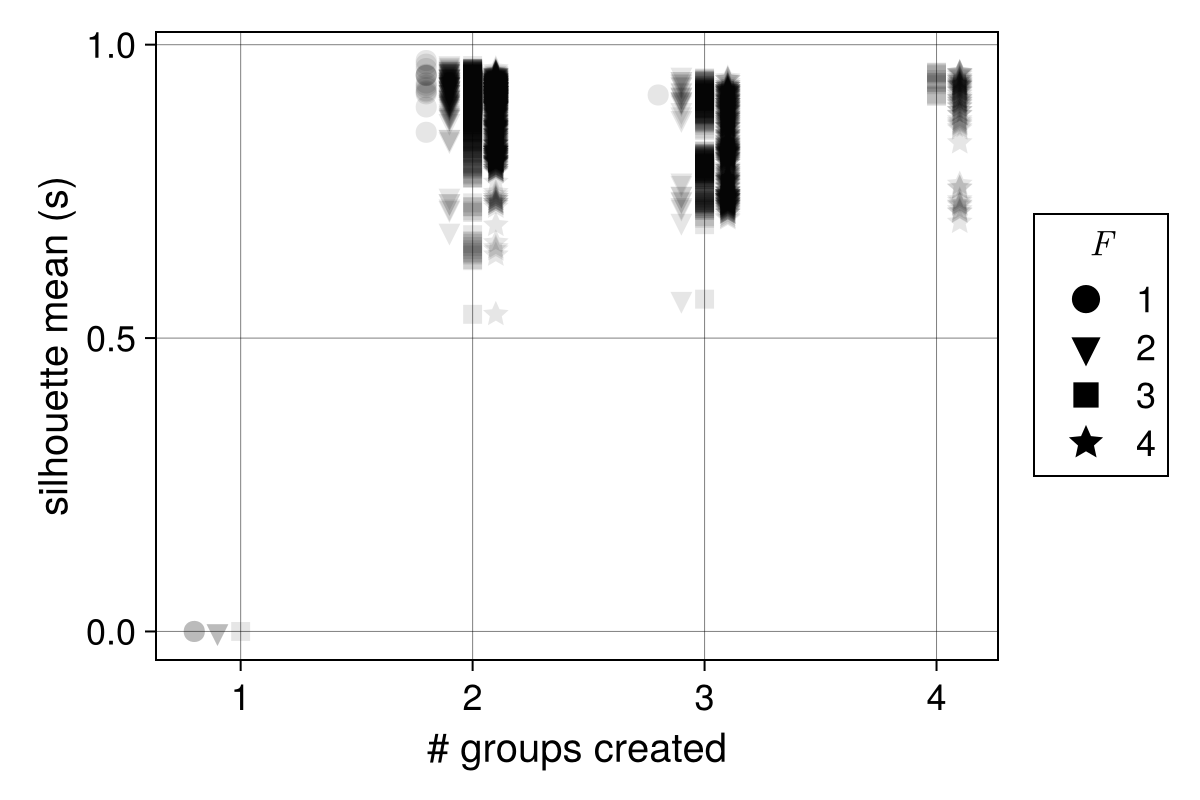}
    \caption{Optimal silhouette mean $s$ during DBSCAN clustering for the Veros data of \S\ref{sec:veros}, versus number of groups created in the clustering. The marker shape corresponds to the number of features used, and markers are plotted with transparency.}
    \label{fig:dbscan_test}
\end{figure}

We now turn on discussing the effect of varying the weights $w_A,w_F$.
The weights are primarily a way so that the practitioner can attribute importance on what they care most: finding more, or less, unique alternative states. As such there is no algorithm to choose ``optimal'' weights.
It is important however that the workflow is overall robust to the chosen weights.
This we demonstrate in Fig.~\ref{fig:sensitivity_of_weights}, where we vary $w_A, w_L$. It is clear that the algorithm output is stable for broad ranges of $w_A, w_F$, as expected, therefore validating the algorithm robustness.
It is important nevertheless for the practitioner to make an informed decision by performing such a sensitivity analysis on their end, as it will likely depend on the exact dataset at hand.

\begin{figure}
    \centering
    \includegraphics[width=0.5\linewidth]{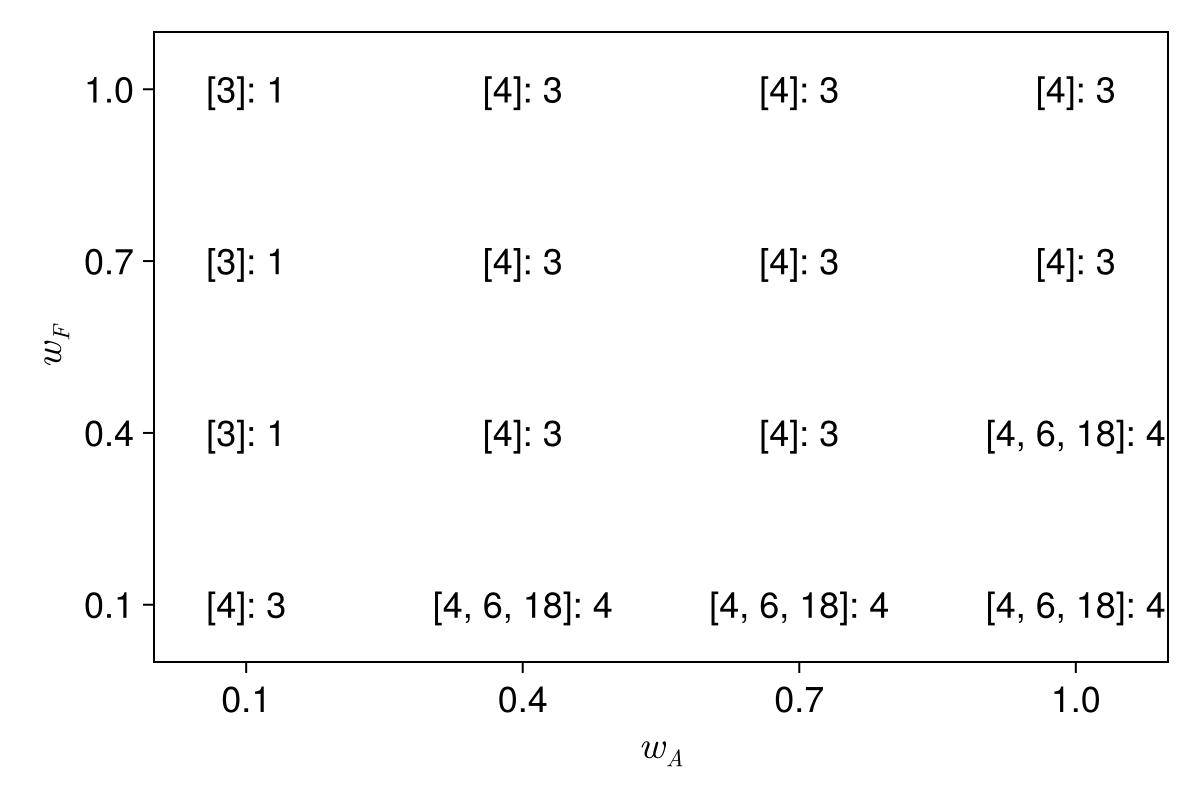}
    \caption{Sensitivity of the workflow output on weights $w_A, w_F$ using the Veros data of \S\ref{sec:veros}. What is shown is the optimal feature dimensions chosen by the workflow, and how unique alternative states they yield: ``[x, y, z]: A''. For performance, we turned off the Iterative option for IA-DBSCAN, so the 4 found attractors for the lower-right part of the plot correspond to the 5 shown in \S\ref{sec:veros}.}
    \label{fig:dbscan_test}
\end{figure}

\section{Appendix C: More on intermingledness}
\label{app:basin_entropy}

Intermingledness was inspired \emph{final state sensitivity}~\cite{Grebogi1983}: the possibility that for some multistable systems, there is an inherent uncertainty for which attractor each initial condition will converge to, because perturbations are just as likely to move the system to another basin than the original basin. This often the case when the basins of individual attractors are strongly mixed with each other (fractal~\cite{McDonald1985} or ``intermingled'').
Measuring this concept is typically done via the basin entropy~\cite{Daza2016} or the uncertainty exponent~\cite{Grebogi1983}. A key disadvantage of both methods (and other related ones) is that they require a dense covering of the basins of attraction. This makes them already inapplicable to our target applications.
Another key problem is that they quantify the basins of attraction as a whole. As such they cannot be applied on a per-group nor on a per-dimension basis, both of which are useful to do for our target applications. In this appendix we compare intermingledness with basin entropy and provide some more results of intermingledness for various types of basins of attraction or data.

\begin{figure}
    \centering
    \includegraphics[width=\linewidth]{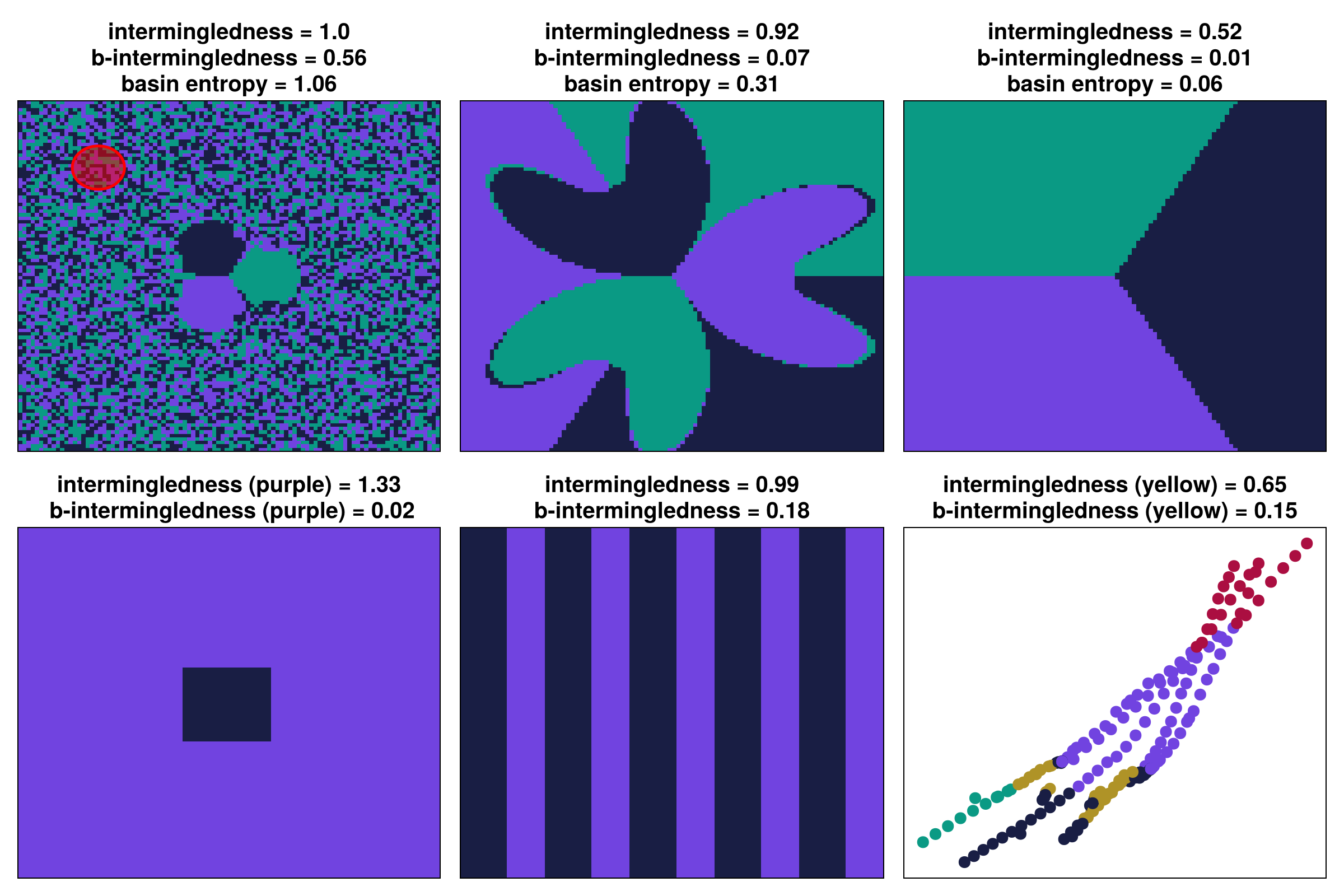}
    \caption{A comparison of intermingledness with basin entropy on paradigmatic basins of attraction. A red circle in the first panel indicates a ball at location $i$ with some radius $r$. The values of intermingledness in the first three panels are all equal between different basins. Intermingledness is computed using the two-dimensional Euclidean distance for $d$.}
    \label{fig:intermingledness}
\end{figure}

Basin entropy is defined as follows. First, we assume that we have access to \emph{densely populated} basins of attraction, as in Fig.~\ref{fig:intermingledness}.
Then, we assume a hypersphere (or hypercube) with size $r$ located at a location $i$ that the basins have been recorded at. Let $p_{ij}$ be the probability that an initial condition in the $i$-th hypersphere will converge to the $j$-th attractor.
Basin entropy is then defined as
$$BE = \frac{1}{N} \sum_{i=1}^{N}\sum_{j=1}^{A} p_{ij}.$$
It quantifies the entropy (uncertainty) of the basins of attraction.

Basin entropy can also be defined only on the boundary set (all $i$ that satisfy $p_{ij} \ne 1 \forall {j}$).
Inspired by this, we devise here an alternative definition of intermingledness, called boundary intermingledness. Following from step 1 of \S\ref{sec:intermingledness} where we have defined the groups $U_a$, we then for each unique pair of groups create the \emph{boundary group} $B_{a,b}$ as follows: first, for all points in $U_a$, find their (unique) nearest neighbours in $U_b$, $N_{a,b}$. Finding nearest neighbours also requires prescribing a distance $d$ so boundary intermingledness can be applied per diagnostic dimension like intermingledness. Now, for all points in $N_{a,b}$ find their (unique) nearest neighbours in $U_a$. The union of these two nearest neighbours sets is $B_{a,b}$. The relative bulk of the boundary set is then $R_{a, b} = |B_{a,b}|/(|U_a| + |U_b|)$.
Boundary intermingledness $BI$ is calculated on a per group basis by averaging the relative bulks with all other groups like normal intermingledness:
$$BI_a^d = \mathrm{summary}\{R_{a,b} \quad \mathrm{for}\quad b \ne a\}.$$
In Fig.~\ref{fig:intermingledness} we compare intermingledness, basin entropy, and boundary intermingledness, in some paradigmatic examples.
We note that all measures considered here, and the uncertainty exponent~\cite{Grebogi1983} which we did not evaluate, are only meaningful within the physical region that has been sampled by the dataset.
They all could change if additional regions further away from the current one are sampled and combined with the current one.

Figure~\ref{fig:intermingledness} makes it clear that intermingledness is not the same as basin entropy, and really quantifies what was originally sought after: the distance needed for a point to be moved to a different basin, relative to the distance needed to stay in its own basin.
Intermingledness can be high because the basins are fractal, or it can be high because the way basins may be positioned.
Although, if the intermingledness of every dimensional projection is $\approx 1$, as in Fig.~\ref{fig:atmos}, this is a clear indication of completely fractalized, likely riddled, basins.
In some very specific cases (such as nested basins, bottom left panel), intermingledness can be misleadingly high.
Unlike basin entropy, and by construction, intermingledness cannot be zero for realistic basins of attraction (that would require completely disconnected basins), but intermingledness can easily be zero when applied to the feature space as in most examples in \S\ref{sec:applications}.

Boundary intermingledness $BI$ on the other hand has more clear connection with basin entropy. It clearly distinguishes between smooth and non-smooth basin boundaries.
On the other hand, it does not really convey whether the basins as a whole are intermingled. For example, in the bottom center panel, one would think that the basins are fully intermingled, as they constantly interlace with each other. But their boundaries are smooth, hence $I = 1$ but $I^{(b)} = 0$.
Ultimately, for real data, the practitioner may decide which quantity to use, or use both. For example, in the Veros dataset shown in bottom right panel, the yellow basin can be considered intermingled when it comes to distance to other basins (in two-dimensional space), or not intermingled when it comes to its smooth boundary.
We also note that $I$ is more robust to adding more data points to the dataset, while $I^{(b)}$ may change substantially depending on where the new data are positioned (bulk or boundary of a basin).

Regardless, intermingledness has some key advantages when compared to traditional quantities used in dynamical systems. (1) it works well for sparse data, (2) its definition is sensible for projections only to specific dimensions (diagnostics or feature dimensions), and (3) it can be calculated for each alternative state individually.

\printbibliography

\end{document}